# Rheological and electrical transitions in carbon nanotube/epoxy suspensions

A. Allaoui and N. El Bounia[*]

IPREM-CANBIO, CNRS UMR 5254, Université de Pau et des Pays de l'Adour, Pau, France

## ABSTRACT

The rheological and electrical properties of suspensions of carbon nanotubes in an uncured epoxy resin were investigated by means of shear rheology and impedance spectroscopy. It was found that above an onset CNT weight fraction (0.1 wt %), the steady viscosity increased with CNT loading and presented a shear thinning behaviour. The concentration dependence of viscosity changed from a power law to an exponential with increasing shear rate, indicating a loss of interaction between aggregates and CNT network breakage. The fluid-to-solidlike and insulator-to-conductor transitions occurred in the same CNT weight fraction range between 0.5 and 0.6 wt %. The correspondence of these transitions was explained by the reduction of contact resistance between CNT by stiffening of the CNT network leading to improved electronic transport.

[*]Corresponding author. Fax: +33 5 59 40 77 44. E-mail address: noureddine.elbounia@univ-pau.fr

## 1. INTRODUCTION

The development of carbon nanotube (CNT) filled-epoxy resin composites has been the subject of an intense research in the last few years. Noticeable improvement of the mechanical and electrical properties was proven [1,2]. The potential applications of these new composites are numerous like high performance materials, conductive adhesives, actuators and sensors. Recently, a great effort was dedicated to the evaluation of the benefits provided by the use of this modified resin as the matrix in conventional fibre reinforced polymer (FRP) composites. The aim was to enhance the mechanical properties through the thickness, that is to say in the transverse direction. The results obtained were far from the expectations, with improvement in the range of 10 to 30% [3-5], due to the uneasy processing and the quite low CNT loading used in these studies. On one hand, improving the dispersion of CNT in a given fluid increases its viscosity. The use of this rheological change as an indicator of the dispersion state of CNT in the fluid was proposed by Huang et al. [6] and recently demonstrated by Cotiuga et al. [7]. On the other hand, the very huge viscosity increase by introduction of CNT in a resin makes the fabric wetting quite laborious. When the resin transfer moulding or infusion or even hand lay-up method is used to fabricate the FRP, filtration and reagglomeration of the CNT by the fabric was reported [3-5]. A good knowledge of the rheological behaviour of the CNT/epoxy suspensions is thus of great interest for the development of these new nano-modified FRP composites. Huang et al. [6] reported on the rheological properties of multiwall nanotubes (MWNT) dispersed in a viscous polymer, showing the relationship between viscosity, mixing time and dispersion state. The experiments were performed at very low gap (0.01 mm) which may not reflect the bulk behaviour. Fan and Advani [8] studied the effect of the preparation method, MWNT aspect ratio, orientation and concentration on the steady and dynamic viscosities of epoxy/MWNT suspensions using a cone and plate rheometer. They reported a fluid-to-solidlike transition at 0.2 wt % of MWNT indicated by the dynamic storage modulus G' plateau at low frequencies. They observed shear-thinning behaviour in steady shear flow. Ma et al. [9] observed the formation of helical bands in CNT/epoxy suspensions in shear flow when the gap was sufficiently small. Rahatekar et al. [10] performed shear flow on CNT/epoxy suspensions along with optical observations. They linked the high viscosity at low shear rates to the presence of interconnected aggregates of CNT and the shear thinning at high shear rates to the breaking of these aggregates, excluding the alignment of CNT as a possible cause. Chen et al. [11] added 1 wt % CNT in epoxy and did not notice significant change on the viscosity due to the inappropriate shear rate range selected for their measurements (most of the points displayed being above 1000 s$^{-1}$). As observed by Rahatekar et al. [10], at these high shear rates, the CNT network is broken and does not contribute significantly to the viscosity. Seyhan et al. [12] compared the dynamic and steady rheological behaviour of CNT/vinyl ester-polyester suspensions with or without CNT amino-functionalisation. They reported frequency-independent storage modulus at low frequencies for CNT loadings above 0.3 wt % and shear thinning behaviour at high shear flow rates. They did not observe any effect of amino-functionalisation on the rheology. Recently Sarvestani [13] proposed for nanoparticle/polymer suspensions a model assuming a strong polymer-particle interaction to explain the solid-like behaviour at low frequency. It was argued that mobility slowing of the polymer chains adsorbed on the nanoparticles is at the origin of this regime. This can be confirmed from literature results [14] where the specific surface of nanoparticles is a key parameter in the viscosity increase. Kim et al. [15] using a capillary rheometer observed a noticeable increase of the elongational viscosity with CNT surface treatments. They attributed it to the improvement of the CNT dispersion and to the interfacial bonding of CNT with the epoxy matrix. Their most efficient treatment led to a three order of magnitude viscosity increase compared to the suspension with untreated CNT. All their suspensions (1 wt% CNT) presented a solid-like behaviour (G' plateau and G'>G''). Kinloch et al. [16] studied the rheological properties of aqueous dispersions of acid-treated nanotubes, they found that the storage modulus scales with CNT concentration as a power law and observed shear thinning behaviour at high shear rates. Hough et al. [17] investigated the dynamic rheological behaviour of aqueous suspensions of surfactant stabilized single wall carbon nanotubes (SWNT). They reported solid-like elasticity evidenced by the frequency independent modulus and the dominance of storage over loss modulus. The storage plateau modulus was found to follow a rigidity percolation law. Xu et al. [18] observed a dependence of the shear thinning behaviour with the length of the nanofiber in steady shear flow for carbon nanofiber (CNF)/glycerol-water suspensions. When the CNF were not chemically treated and thus sufficiently long, the viscosity increases and the shear thinning was high. In the contrary, there were almost no shear thinning and little viscosity increase when short chemically treated and sonicated CNF were used to prepare the suspensions. The authors also reported elastic solid like behaviour as indicated by a storage modulus low-frequency plateau for sufficiently high CNF loadings (> 3 wt %). Kotsilkova et al. [19] found a scaling law for storage modulus with volume fraction in the case of carbon black nanoparticles dispersed in epoxy resin. They evidenced two regimes differing by the exponent of the scaling law, the crossover volume fraction being taken as a critical value separating a weak-interaction from a stronger interaction domain between the nanoparticles.

This short review of the literature shows that most of the features of the rheological behaviour of carbon nanotube suspensions were observed: the shear thinning, the fluid-to-solid like transition, the viscosity increase with CNT loading, the dependence of viscosity with the dispersion state, the orientation and the aspect ratio of CNT.

In this article, the steady shear flow and dynamic rheology of CNT dispersed in an uncured epoxy resin were investigated in order to address the CNT concentration dependence of the viscosity and investigate more deeply the rheological transitions in CNT suspensions. Additionally, the electrical properties were measured by impedance spectroscopy in order to address the relation between rheological and electrical transitions.

## 2. EXPERIMENTAL

### 2.1 Materials

Multiwall carbon nanotubes MWNT (Graphistrength™ with diameters in the range of 10-15 nm and lengths between 0.1 and 10 μm, were supplied by Arkema. A diglycidyl ether of bisphenol-A (DGEBA) epoxy resin (LY556, Araldite) was used.

### 2.2 Sample preparation

A masterbatch of 1.5 wt % MWNT in epoxy was prepared by high shear mixing with a Silverson at 1200 RPM during 30 minutes followed by mixing with a colloidal mill during 30 minutes at 12 m/s (3400 RPM). Dilutions of the masterbatch (from 1.37 to 0.00012 wt % MWNT) were obtained by adding a given amount of epoxy and homogenizing the sample with a high speed mixer (a drill equipped with a home-made paddle) at 3000 RPM during 5 minutes.

**2.3 Rheology and impedance spectroscopy**

Rheological measurements in shear flow mode were performed with a Rheometrics DSR stress-controlled rheometer in plate–plate geometry (diameter of plates 25 mm) at room temperature with a gap in the range of 1 to 1.8 mm. A sequence of 4 shear rate sweeps alternatively at increasing and decreasing shear rate (between 0.01 and 100 s$^{-1}$) was applied to each sample in order to erase the mechanical history and structuration effects. Frequency sweeps from 10 down to 10$^{-4}$ rad/s were performed with the DSR in the linear viscoelastic domain at a constant stress of 5 Pa. The electrical properties were investigated using a HP 4192A impedance analyzer equipped with a coaxial cylindrical liquid cell (Ferisol CS601). All experiments were performed on fresh samples (before rheology) at room temperature and testing frequencies ranged from 10$^3$ to 10$^7$ Hz. The AC conductivity is given by

$$\sigma_{AC} = \frac{(S - S_0)\varepsilon_0}{C_a}$$

with $S$ and $S_0$ being the conductance of the cell respectively with and without sample, $\varepsilon_0$ the permittivity of free space and $C_a$ the active capacity.

**3. RESULTS AND DISCUSSION**

**3.1 Steady shear viscosity**

For all samples, only the last flow of a sequence of 4 shear rate sweeps alternatively at increasing and decreasing shear rate $\dot{\gamma}$ was analyzed. The shear flow viscosity as a function of shear rate is plotted in Fig. (**1**)a. At low CNT loadings, the samples were similar to the unfilled resin with a Newtonian behaviour and a viscosity around 10-12 Pa.s. Above 0.148 wt % CNT, the samples start exhibiting a shear thinning behaviour with a substantial increase of the viscosity compared to the neat resin. The low shear rate viscosity $\eta$ increases with the CNT loading and follows a power law behaviour: $\eta \alpha \dot{\gamma}^n$. The viscosity increase was much less consequent at high shear rates due to shear thinning. As depicted in Fig. (**1**)b, the exponent $n$ (pseudoplastic index) was found to increase with the CNT loading until an asymptotic value of ~ 0.8. The ($n$, CNT wt%) curve is well fitted with an exponential function, $n = n_0 - A\exp(Bp)$ where $n_0$, $A$ and $B$ are constants and $p$ is the CNT weight fraction. It is worth noting that the CNT concentration dependence of the viscosity was exponential at high shear rates and followed a power law at low shear rates as shown in Fig. (**2**). This concentration dependence change, which occurred between 0.1 and 0.5 s$^{-1}$, could be related to the loss of interaction between aggregates at high shear rates and the breakage of the CNT network. To our knowledge, this is the first observation of this phenomenon which can be considered as a rheological transition. The viscosity increase and the concentration dependence change occurred above an onset CNT loading around 0.1 wt %. An estimation of the overlap volume fraction, the minimum fraction to form a connected network, is provided by the rigid rod theory $\phi = 3/2(\ell/d)^{-2}$ with $\ell$ and $d$ the length and diameter of the rod. A CNT weight fraction of 0.1 wt % corresponds to an aspect ratio $\ell/d \sim 55$ in good agreement with the dimensions of the CNT (diameters 10-15 nm, average length 1 µm) used in the present study.

## 3.1 Dynamic viscosity and conductivity

The dynamic measurements revealed a transition from a fluid to a solid-like behaviour between 0.5 and 0.6 wt % as indicated by the storage modulus plateau at low frequency (Fig. (**3**)). The low frequency storage modulus which represents the elastic character is plotted in Fig. (**4**) as a function of CNT wt %. It was found that the data could be fitted to the rigidity percolation law with an exponent around 4.2 and a critical weight fraction of 0.1 wt %. A critical exponent of 2.3 was reported by Hough et al. [17] for aqueous suspensions of surfactant stabilized SWNT. Sahimi et al. [20] performed bond percolation simulations of three-dimensional elastic networks and predicted a critical exponent of 2.1 ± 0.2 for bonds resisting stretching and free to rotate, and a value of 3.75 ± 0.11 in the case of bonds resisting both stretching and rotation. We found a higher value, however, it should be noted that we may not have enough data points close to $p_c$ for a reliable fit. In Fig. (**4**), the electrical conductivity at 1 Hz obtained by extrapolation of experimental data was plotted. It is worth noting that the insulator-to-conductor transition occurred in the same range of CNT weight fraction. This transition was indicated by the low frequency plateau of the conductivity (Fig. (**5**)). Du et al. [21] reported a rheological threshold smaller than the electrical percolation threshold for CNT/PMMA nanocomposites. They explained their result by the smaller CNT-CNT distance required for conduction compared to that required to limit polymer chain mobility when considering the radius of gyration of the polymer chains. It should be noted that in their study the rheological threshold was obtained from measurements in the melt state at 200°C while the electrical percolation threshold was deduced from room temperature measurements on the same samples in the solid state. Moreover, the threshold was obtained by fitting a few points and may not be precise. In our case, the rheological and electrical measurements were performed at room temperature in the liquid state. The transitions were determined from the change in the frequency behaviour (from frequency dependent to independent). For our results, the correspondence of rheological and electrical transitions is explained by the small radius of gyration of the epoxy chains and by the conditions of efficient electronic transport through the suspensions. The formation of a CNT network (connectivity) spanning the sample is the first condition. Nevertheless, the electronic transport is governed by the quality of contacts between CNT inside this network. In the fluid to solid-like transition, the network stiffens and the contacts between CNT are better and also stable so that the electronic transport is also improved by reduction of contact resistance between CNT (with or without tunnelling through the insulating layer between CNT). A contact network model based on these principles was previously shown to satisfactory describe the conductivity of carbon nanofibre/epoxy composites [22].

## 4. CONCLUSIONS

In this study, the rheological and electrical properties of a masterbatch of carbon nanotubes dispersed in an uncured epoxy resin and its dilution were investigated by steady and dynamic shear rheology and impedance spectroscopy. Above an onset CNT loading around 0.1 wt %, the viscosity increased substantially with the concentration and the suspensions presented a shear thinning behaviour. The CNT concentration dependence of the viscosity was found to follow a power law form at low shear rate and an exponential law at high shear rates. The concentration dependence change was related to the loss of interaction between aggregates and the breakage of the CNT network at high shear rates. In the dynamic measurements, the fluid-to-solidlike and the insulator-to-conductor transitions were found to be in the same range between 0.5 and 0.6 wt %. The correspondence of the two transitions was explained by the stiffening of the CNT network leading to improved contacts and lower contact resistance between CNT. The efficiency of the electronic transport through the CNT network is linked to the network rigidity due to the contact resistance between CNT.

## ACKNOWLEDGEMENTS

Pr. J-P. Ibar is greatly acknowledged for fruitful discussions. Financial support from ARKEMA is appreciated.

**LIST OF CAPTIONS**

Figure 1 – a) Shear flow viscosity as a function of shear rate and b) low shear rate power law behaviour exponent as a function of CNT wt%.

Figure 2 – Steady shear viscosity as a function of CNT wt % at different shear rates.

Figure 3 – Storage modulus as a function of frequency for different CNT loadings.

Figure 4 – Electrical conductivity and storage modulus at 0.1 and 0.01 rad/s as a function of CNT wt % (inset: fit of storage modulus to rigidity percolation law).

Figure 5 – Electrical conductivity as a function of frequency for different CNT loadings.

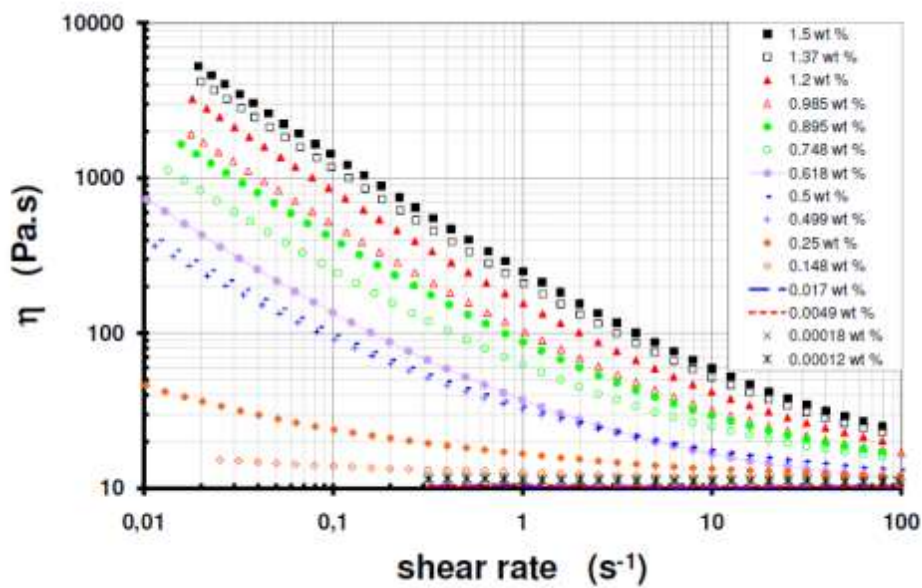

**Figure 1 – a) Shear flow viscosity as a function of shear rate.**

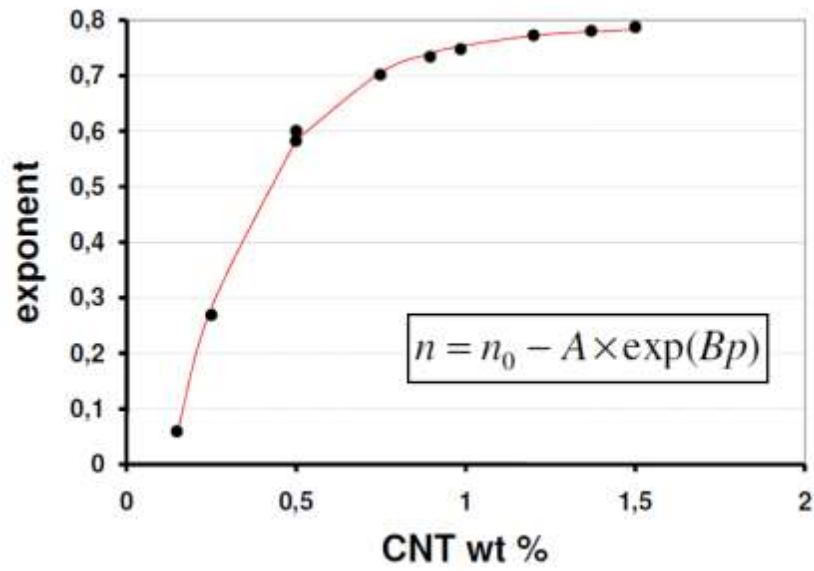

**Figure 1 – b) low shear rate power law behaviour exponent as a function of CNT wt%.**

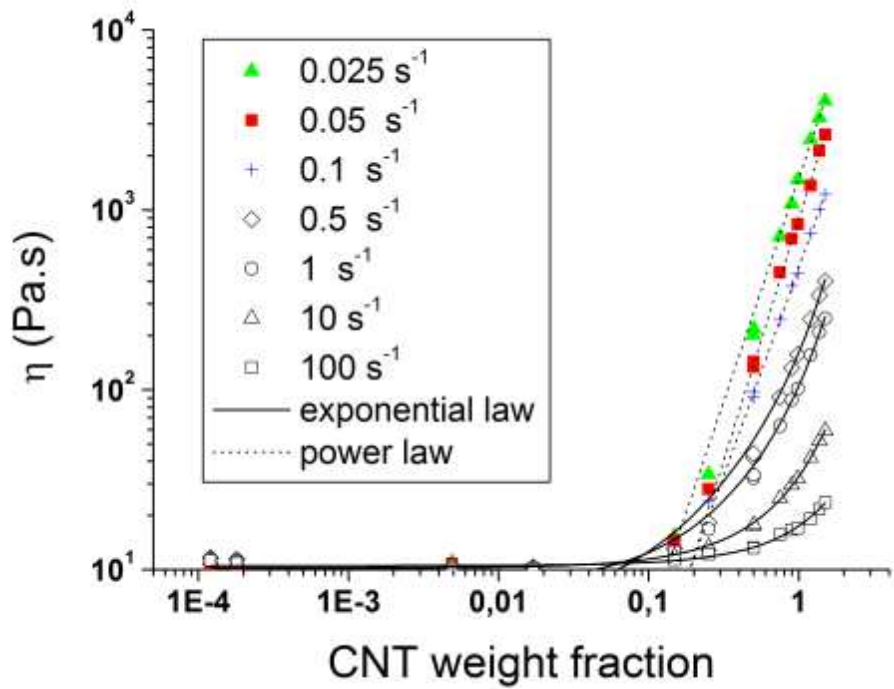

Figure 2 – Steady shear viscosity as a function of CNT wt % at different shear rates.

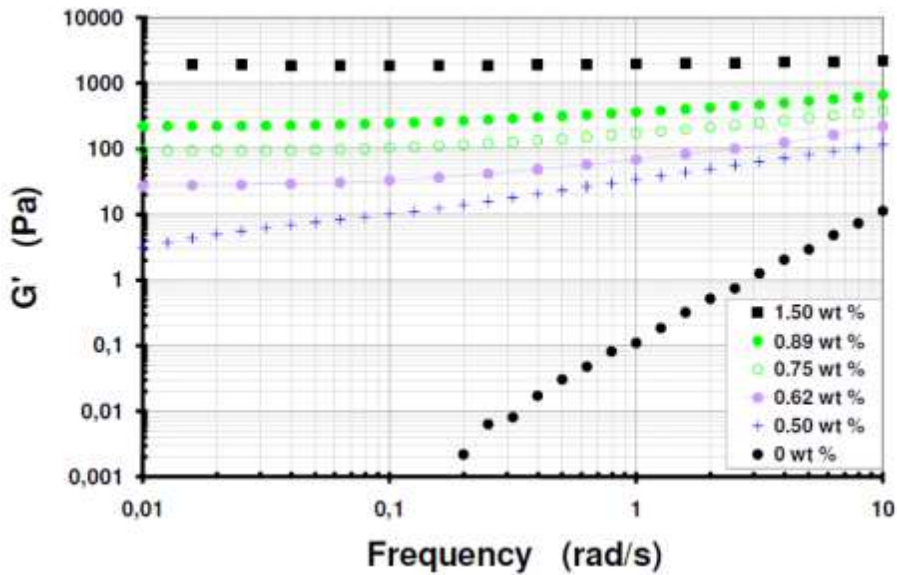

Figure 3 – Storage modulus as a function of frequency for different CNT loadings.

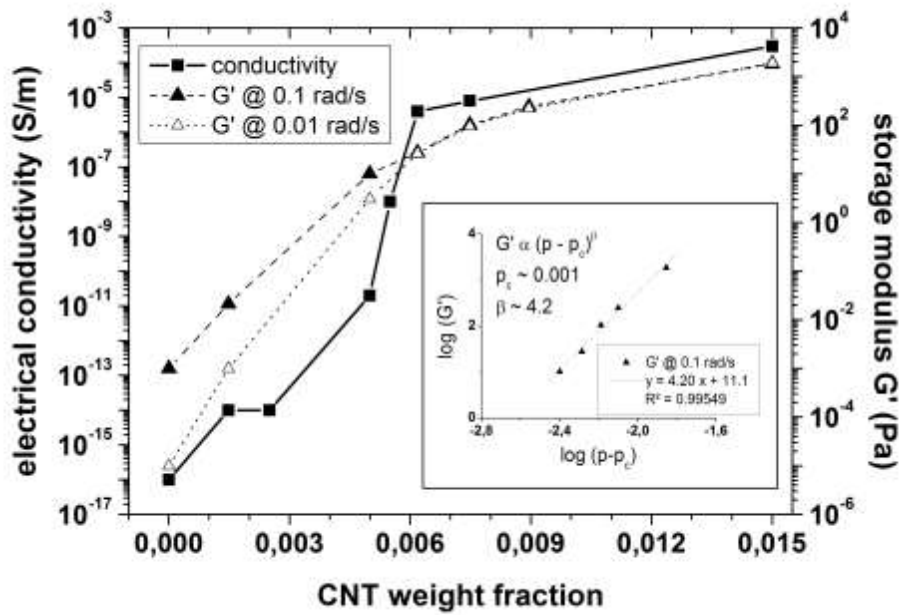

Figure 4 – Electrical conductivity and storage modulus at 0.1 and 0.01 rad/s as a function of CNT wt % (inset: fit of storage modulus to rigidity percolation law).

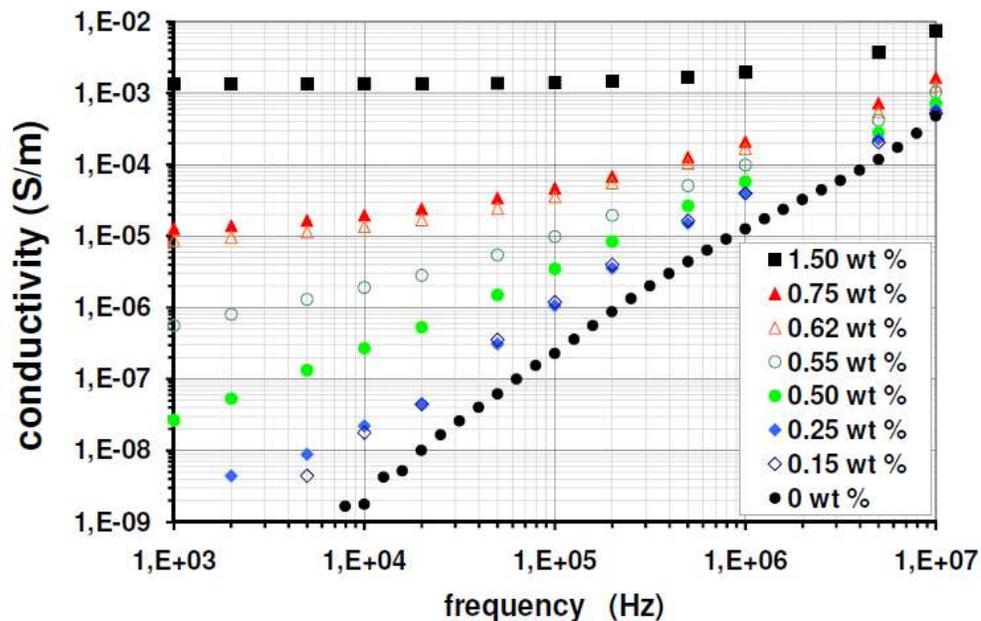

**Figure 5 – Electrical conductivity as a function of frequency for different CNT loadings.**